\documentclass[twocolumn,showpacs,preprintnumbers,amsmath,amssymb]{revtex4}

\usepackage{graphicx}
\usepackage{dcolumn}
\usepackage{bm}

\begin{document}

\title{Holographic Gravity and the Surface term in the Einstein-Hilbert Action}

\author{T. Padmanabhan}
\email{nabhan@iucaa.ernet.in}

\affiliation{IUCAA, Pune University Campus,\\
P.B. 4, Ganeshkhind, Pune 411 007, INDIA.}

\date{\today}

\begin{abstract}
Certain peculiar features of Einstein-Hilbert (EH) action provide  clues towards
a holographic approach to gravity  which is 
independent of the detailed  microstructure of spacetime. 
These features of the EH action include:
 (a) the existence of second derivatives of dynamical variables;
(b) a non trivial relation between the surface term and the bulk term;
(c) the fact that  surface term is non analytic in the coupling constant, when gravity is treated
as a spin-2 perturbation around flat spacetime 
and (d) the form of the  variation of the surface term under infinitesimal coordinate transformations. The surface term can be derived directly from very general considerations and using
(d) one can obtain Einstein's equations {\it just from the surface term of the action}.
Further one can relate the bulk term to the surface term and derive the full
 EH action based on purely thermodynamic considerations. The features
 (a), (b) and (c) above emerge in a natural fashion in this approach. It is shown that action $A_{grav}$ splits into two
 terms $-S+\beta E$ in a natural manner \textit{in any stationary
 spacetime with horizon}, where $E$ is essentially an integral over ADM energy density and $S$ arises from the integral
 of the surface gravity over the horizon.
 This analysis shows that the true degrees of freedom of gravity reside
in the surface term of the action, making gravity intrinsically holographic. It also provides a close connection
between gravity and gauge theories, and highlights the subtle role of 
the singular coordinate transformations.
\end{abstract}

\maketitle

\section{Introduction}

If we treat the macroscopic spacetime as analogous to a continuum solid and the unknown
microscopic structure of spacetime as analogous to the atomic structure \cite{sakharov}, then it is possible
to gain some important insights into the possible nature of quantum gravity. First of all, we note that the macroscopic description 
of a solid uses concepts like density, stress and strain, bulk velocity etc., none
of which can even be usefully defined in the microscopic description. Similarly,
variables like metric tensor etc. may not have any relevance in quantum gravity.
Second,
the quantum theory of a spin-2 field (``graviton'') will be as irrelevant in quantum gravity,
as the theory of phonons in providing any insight into the electronic structure of atoms.
Third, the symmetries of the continuum description (e.g., translation, rotation etc.) will
be invalid or will get strongly modified in the  microscopic description. A naive
insistence of diffeomorphism invariance in the quantum gravity, based on the classical
symmetries, will be as misleading as  insisting on infinitesimal rotational invariance of, say,
an atomic crystal lattice. In short, 
the variables and the description will change in  an  (as yet unknown)  manner.
It is worth remembering that the Planck scale ($10^{19}$ GeV) is much farther away
from the highest energy scale we have in the lab ($10^2$ GeV) than the atomic
scale  ($10^{-8}$ cm) was from the scales of continuum physics (1 cm).

It is therefore worthwhile to investigate  general features of quantum microstructure
which could be reasonably independent of the detailed theory of quantum spacetime ---
whatever it may be. I will call this  a   \textit{thermodynamic} approach to spacetime dynamics,
to be distinguished from the statistical mechanics of microscopic spacetimes \cite{tppr}. To do this,
I will exploit the well known connection between thermodynamics and the physics of
horizons \cite{gen1,gen2} but will turn it on its head to \textit{derive} the Einstein's equations and the Einstein-Hilbert 
action from thermodynamic considerations \cite{pr154}. This procedure will throw light on several
peculiar features of gravity (which have no explanation in the conventional approach)
and will provide a new insight in interpreting general coordinate transformations.

\section{Observers and their Horizons}

Principle of Equivalence, combined with special relativity, implies that gravity will affect the 
trajectories of light rays and hence the causal relationship between events in spacetime. In particular, there will exist families of observers (congruence of timelike curves) in any spacetime
who will have access to only part of the spacetime. 
Let a timelike curve $X^a(t)$, parametrized by the proper time $t$ of the clock
   moving along that curve, be the trajectory of an observer in such a congruence and let $\mathcal{C}(t)$
   be the past light cone for the event $\mathcal{P}[X^a(t)]$ on this trajectory. The union $U$ of all these past light cones 
   $\{\mathcal{C}(t),-\infty\leq t \leq \infty\}$ determines whether an observer on the trajectory $X^a(t)$
   can receive information from all events in the spacetime or not. If $U$ has a nontrivial boundary, there will be regions in the spacetime from which this observer cannot receive signals. 
The boundary of the union of causal pasts of \textit{all} the observers in the congruence --- which is essentially  the boundary of the union of backward
light cones ---  will define a {\it causal} horizon for this congruence. (In the literature, there exist different definitions
for horizons appropriate for different contexts; see e.g.\cite{pr30}; we will use the above definition.) 
This horizon is
  \textit{dependent} on the family of observers that is chosen, but is \textit{coordinate independent}.
  
 A general class of metrics with such a  static horizon can be described 
 \cite{comment0} by the line element
 \begin{equation}
   ds^2=-N^2 (x^\alpha)  dt^2 +  \gamma_{\alpha\beta} (x^\alpha) dx^\alpha dx^\beta
   \label{startmetric}
   \end{equation}
   with the conditions that:
   (i) $g_{00}({\bf x}) \equiv -N^2({\bf x})$ vanishes on some 2-surface $\mathcal{H}$;
   (ii) $\partial_\alpha N$ is finite and non zero on $\mathcal{H}$
   and (iii) all other metric components and curvature
   remain finite and regular on $\mathcal{H}$. The natural congruence of observers with ${\bf x}=$
    constant will perceive $\mathcal{H}$ as a horizon. The four-velocity $u_a=-N\delta^0_a$ of
    these observers has a corresponding  four acceleration $a^i=
    u^j\nabla_ju^i=(0,{\bf a})$ with $a_\alpha=(\partial_\alpha N)/N$.
 If $n_a$ is the unit normal  to the $N=$ constant surface, then
    the `redshifted' normal component of the acceleration $N(a^i n_i)  = ( g^{\alpha\beta} \partial_\alpha N \partial_\beta N)^{1/2}\equiv Na({\bf x})$
(where the last equation defines the function $a$) has a finite limit on the horizon. On the horizon $N=0$, we take
$Na\to \kappa$ where $\kappa$ is called the surface gravity of the horizon (see e.g., \cite{pr153}). 
(The results extend to stationary spacetimes but we will not discuss them here.)

These static spacetimes have a more natural  coordinate system defined {\it locally} in terms of the level surfaces of $N$. That is, we transform from the original space coordinates $x^\mu$ in Eq.(\ref{startmetric}) to the set $(N,y^A), A=2,3$ (where $y^A$ are transverse coordinates on the
$N=$ constant surface) by treating $N$ as one of the  spatial coordinates. The metric can now be transformed to the form 
\begin{equation}
ds^2=-N^2dt^2+ \frac{dN^2}{(Na)^{2}}+
dL_\perp^2
\label{iso}
\end{equation}
where $dL_\perp^2$ is the metric on the transverse plane which is relatively unimportant for our discussion.
Near the $N\to 0$ surface, $Na\to \kappa$ and the metric reduces to the (Rindler) form:
   \begin{equation}
   ds^2
   \simeq-N^2 dt^2 + \frac{dN^2}{\kappa^2} +dL_\perp^2
   =-\kappa^2x^2 dt^2 + dx^2 +dL_\perp^2
   \label{dsfirst}
   \end{equation}
   with $x=N/\kappa.$ This (Rindler) metric 
    is a good approximation to a large class of static metrics with $g_{00}$ vanishing on a surface
    which we have set at $N=0$.

In \textit{classical} theory,  the horizon at $N=0$  acts as a one-way membrane and shields the observers at $N>0$ from the processes that take place on the `other side' of the horizon ($N<0)$.
However, this is no longer true in \textit{quantum} theory since entanglement and tunneling across the horizon can lead to nontrivial effects \cite{pr10}. This is obvious in the study of quantum field theory in a spacetime partitioned by a horizon. The two point function in quantum theory is non zero for events separated by spacelike intervals leading to non-zero correlations. Or, rather, quantum field theory can only be formulated in the Euclidean sector of the spacetime (or with an $i\epsilon$ prescription, which is the same thing) and the Euclidean sector contains information from across the horizon. Hence the causal partitioning of spacetime by a horizon --- which is impenetrable in classical theory --- becomes porous in quantum theory. 

Nevertheless, it seems reasonable to postulate that \textit{any} class of observers have a right to describe physical phenomena entirely in terms of the variables defined in the regions accessible to them \cite{pr148}.  Mathematically, this will require using a coordinate system in an \textit{effective Euclidean} manifold in which the inaccessible region is removed \cite{pr57}. 
Near any static horizon one can set up the Rindler coordinates in Eq.(\ref{dsfirst}) which has the Euclidean extension (with $\tau=it$): 
\begin{equation}
ds^2_E\approx N^2 d\tau^2 +dN^2/\kappa^2+dL_\perp^2
\label{eext}
\end{equation} 
This covers the region outside the horizon ($N>0$) with the horizon mapping to the origin; removing the region inside the horizon is equivalent to removing the origin from the $\tau-N$ plane.
Any nontrivial quantum effect due to horizon should still have a natural interpretation in this effective manifold and indeed it does. The effective Euclidean manifold
acquires a nontrivial topology and the standard results (like the thermal effects) of quantum field theory in curved spacetime arises from this nontrivial topology \cite{pr57,pr59}. 

\section{A holographic derivation of Einstein's equations}

In the present work, we are more interested in exploring the consequences for gravity itself which we shall now describe. 
Since horizon has  forced us to remove a region from the manifold, we are also forced to deal with manifolds with non trivial boundaries, both in the Euclidean and Lorentzian sectors. (In the Lorentzian sector we shall approach the horizon as a limit of a sequence of timelike surfaces
e.g., we take $r=2M+\epsilon$ with $\epsilon\to +0$ in the Schwarzschild spacetime). The action
functional describing gravity {\it will} now depend on variables defined on the boundary of this region. Since the horizon (and associated boundaries) may exist for some observers (e.g., uniformly accelerated observers in flat spacetime, $r=$ constant $>2M$ observers in the Schwarzschild spacetime ...) but not for others (e.g, inertial observers in flat spacetime, freely falling observers
inside the event horizon, $r<2M$, in the Schwarzschild spacetime ), this brings up a new level of observer dependence in the theory. It must, however, be stressed that this view point is completely in concordance with what we do in other branches of physics, while defining action functionals. The action describing QED at $10$ MeV, say,
does not use degrees of freedom relevant at $10^{19}$ GeV which we have no access to. Similarly, if an observer has no access to part of spacetime, (s)he should be able to use an action principle
using the variables (s)he can access, which is essentially the philosophy of renormalisation group theory translated into real space
from momentum space \cite{pr148}. This brings about the boundary dependence in the presence of horizons.
 
Further, since we would like the action to be an integral over a local density, the surface term must arise from integrating a four-divergence term in the Lagrangian and the gravitational action functional (in the \textit{Euclidean sector}, which we shall consider first) will have a generic form:
\begin{equation}
A_{\rm grav}  = \int_{\cal V} d^4 x \sqrt{g} \, \left( L_{\rm bulk}+ \nabla_i U^i \right)
= A_{\rm bulk} + A_{\rm sur}
\label{firsteqn}
\end{equation}
The vector $U^a$ has to be built out of the normal $u^i$ to the boundary $\partial{\cal V}$ of ${\cal V}$, metric $g_{ab}$ and
the covariant derivative operator $\nabla_j$ acting at most  once. The last restriction arises because the equations of motion
should be of no order higher than two. (The normal $u^i$ is defined only on the boundary $\partial\mathcal{V}$ but we can extend it to the bulk $\mathcal{V}$, forming a vector field, in any manner we like since the action only depends on its value on the boundary.)
 Given these conditions,
there are only four possible choices for $U^i$, viz.
$(u^j\nabla^i u_j, u^j\nabla_j u^i, u^i\nabla^j u_j,u^i)$. Of these four, the first one identically vanishes since $u^j$ has unit norm; the second one --- which is
the acceleration $a^i=u^j\nabla_j u^i$ vanishes on integration since the boundary term is $u_iU^i=a^iu_i=0$. Hence the most general vector $U^i$ we need to consider is the linear combination of $u^i$ and $Ku^i$ where $K\equiv -\nabla_iu^i$ is the trace of the extrinsic curvature of the boundary. Of these two,  $U^i=u^i$ will lead to the volume of the bounding surface which we will ignore. (It can be, in general, divergent and hence is not an acceptable candidate. In any case, retaining it does not alter any of our conclusions below). Thus the  surface term (arising from $Ku^i$) must have the form 
\begin{equation}
\label{threedaction}
A_{\rm sur}\propto\int_\mathcal{V} d^4 x \, \sqrt{g} \nabla_i (Ku^i)
=
\frac{1}{8\pi G}\int_{\partial\mathcal{V}} d^3 x \, \sqrt{h} K 
\end{equation}
where $G$ is a  constant to be determined (which has the dimensions of area in natural units
with $c=\hbar=1$) and  $8\pi$ factor is introduced with some hindsight. 

What does the surface term contribute on the horizon?   Consider a surface 
$N=\epsilon, 0<\tau<2\pi/\kappa$ and the full
range for the transverse coordinates; this surface 
is infinitesimally away from the horizon in the \textit{Euclidean} spacetime described by Eq.(\ref{eext}) and has the unit normal $u^a=\kappa(0,1,0,0)$. Its contribution to the action is the integral of $K=-\nabla_a u^a=-(\kappa/\epsilon)$ over the surface:
\begin{equation}
A_{\rm sur}=-\frac{1}{8\pi G}\int d^2 x_\perp \int_0^{2\pi/\kappa}d\tau 
\epsilon \left( \frac{\kappa}{\epsilon}\right) =-\frac{1}{4}\frac{\mathcal{A}_\perp}{G}
\label{seven}
\end{equation}  
which is (minus) one quarter of the transverse area $\mathcal{A}_\perp$ of the horizon, in units of $G$ (which is still an undetermined constant). This contribution is universal and of course independent of $\epsilon$ so that the limit
$\epsilon\to0$ is trivial. (Also note that the term we ignored earlier, the integral over the four-volume of  $\nabla_au^a$, will not contribute on a horizon where $\epsilon\to 0$). Since the surface contribution is due to removing the inaccessible region, it makes sense to identify $-A_{\rm sur}$ with an entropy. The sign in Eq.(\ref{threedaction}) is correct with $G>0$ since we expect --- in the Euclidean sector --- the relation $\exp(-A_{\rm Euclid})=\exp S$ to hold, where $S$ is the entropy.

Analytically continuing to the Lorentzian sector it is possible to show that (see Appendix A of ref.\cite{tppr}) the surface term gives the contribution
\begin{equation}
A_{\rm sur}=-\frac{1}{16\pi G}\int_\mathcal{V} d^4 x\partial_a P^a
\label{sur}
\end{equation}
where
\begin{equation}
    P^a  =-
     \frac{1}{\sqrt{-g}} \partial_b (g g^{ba})=
    \sqrt{-g} \left(g^{ak} \Gamma^m_{km}- g^{ik} \Gamma^a_{ik}\right)
    \label{defpcone}
     \end{equation}
It is  clear that, while $L_{\rm bulk}(g,\partial g)$ in Eq.(\ref{firsteqn}) can be made to vanish at any given event by going to 
the local inertial frame (in which $g=\eta,\partial g=0$), one cannot make $\partial_a P^a$ vanish in a local inertial
frame. In such a frame, we have $\partial_a P^a = -\partial_a \partial_b g^{ab}$.
This suggests that the true dynamical degrees of freedom of gravity reside in the surface
term rather than in the bulk term, making gravity intrinsically holographic. Obviously,
 the really important term in the Hilbert action
is the often neglected surface term! To understand this term which leads to the entropy of 
the horizon, let us explore it in a few simple contexts. 

To begin with,
consider spacetime metrics of the form $g_{ab} = \eta_{ab} + h_{ab}$ with $h_{ab}$ treated
as a first order perturbation. In this case, $P^a = \partial_b(\eta^{ab} h^i_i - h^{ab})$
and the surface term
\begin{equation}
A_{\rm sur}=-\frac{1}{16\pi G}\int_\mathcal{V} d^4 x\partial_a \partial_b(\eta^{ab} h^i_i - h^{ab})
\label{sur1}
\end{equation}
 is gauge invariant under the transformations $h_{ab}\to h_{ab}+\partial_a\xi_b+\partial_b\xi_a$.
(Sometimes it is claimed in the literature that a term which is invariant under such infinitesimal gauge
transformations will be generally covariant under finite transformations. This is clearly not true and $A_{\rm sur}$
is an instructive counter example.)
In the Newtonian limit with $g_{00} = - (1+2\phi)$, this leads to ${\bf P} = 2\nabla \phi=-2\textbf{g}$
which is proportional to the gravitational acceleration. The contribution from any surface
is then clearly the normal component of the acceleration, ie., surface gravity, \textit{even in the Newtonian limit.} [As an aside, let me mention that this
 is rather intriguing. It is known that thermal effects of horizons have a classical analogue (see e.g.,\cite{srini})
 and that statistical mechanics of systems with Newtonian gravity
  has several peculiar features (see e.g.,\cite{sm}); but one rarely studies matter interacting by
  Newtonian gravity from a (limit of the) action functional. It is not clear whether this result is of any deep
  significance.]

Let us next consider how the surface contribution varies when one goes from a local inertial frame
to an accelerated frame by an infinitesimal transformation $x^a\to x^a+ \xi^a$ \textit{in flat spacetime}.
 It is easy to show that, in this case, $P^a=-\partial_i(\partial^a \xi^i - \partial^i \xi^a)$.
 (While the variation in the metric depends on the 
 symmetric combination $\delta g^{ai}=\partial^a \xi^i + \partial^i \xi^a$, the contribution to $P^a$
 arises from the antisymmetric combination.) For regular (non-distributional) functions, $\partial_aP^a $ vanishes
 showing that the contribution from any given surface $P^a n_a$ is a constant. We are interested in
 the  infinitesimal version of the transformation from inertial to Rindler coordinates, which 
 corresponds to $\xi^t = -\kappa t x, \xi^x = -(1/2)\kappa t^2$. The surface
 term now picks up the contribution $2 \kappa$ on each surface. 
 Thus the surface term has a purely local interpretation and is directly 
 connected with the acceleration measured by local Rindler observers. ( When $P^a$ is a constant
 $\partial_aP^a$ will vanish but one can still work out the constant contribution $n_aP^a$ from a surface with normal $n_a$ and interpret its value. The contribution we computed in Eq.(\ref{seven}) is indeed from the Rindler frame contribution \textit{but evaluated on a surface}.)

 Since $(-A_{sur})$ represents the entropy, its variation has direct thermodynamic significance. To obtain gravity by a thermodynamic route,
we will take the      
 total action $A_{tot}$ for matter plus gravity to be the sum of $(-A_{sur})$ and the standard matter action $A_{matter}[\phi_i,g]$ in a spacetime with metric
$g_{ab}$. The $\phi_i$ denotes some matter degrees of freedom, the exact form of which does not concern us; varying $\phi_i$
will lead to standard equations of motion for matter in a background metric and these equations will also ensure that the energy momentum tensor
of matter $T^a_b$ satifies $\nabla_aT^a_b=0$.

We will now prove the key result of this section: Einstein's equations arise 
from the demand that $A_{tot}=-A_{sur}+A_{matter}$  should be invariant under virtual displacements of the horizon normal to itself.

Let $\mathcal{V}$ be a region of spacetime such that part of the boundary of the spacetime $\partial\mathcal{V}$
is made up of the horizon $\mathcal{H}$. [For example, in the Schwarschild metric we can take $\mathcal{V}$ to be bounded by the surfaces $t=t_1,t=t_2,r=2M,r=R>2M$.]
Consider
an infinitesimal coordinate transformation
$x^a\to \bar{x}^a=x^a+\xi^a$, where $\xi^a$ is nonzero only on the horizon and is in the direction of the normal to the horizon --- which makes it a null vector. Clearly, one can think of this transformation as making a virtual displacement of horizon normal to itself. Under $x^a\to \bar{x}^a=x^a+\xi^a$, the metric changes by
$\delta g^{ab}=\nabla^a\xi^b+\nabla^b\xi^a$ and the matter action changes by $(\delta A_{matt}/\delta g^{ab})=-(1/2)\sqrt{-g}T_{ab}$. Using
$\nabla_a T^a_b=0$, this can be written as:
\begin{equation}
\delta A_{matt}=
=-\int_\mathcal{V}d^4x\sqrt{-g}\nabla_a(T^a_b\xi^b)
\label{delmat}
\end{equation}
 Next, to find the explicit form of   $(\delta A_{sur}/\delta g^{ab})$ under infinitesimal coordinate transformations, we can either work explicitly with Eq.(\ref{sur}) or use the fact that the variation of the surface term arises from the
integration over $g^{ab}\delta R_{ab}$ in the action\cite{comment3}. This gives  
\begin{equation}
\delta(- A_{\rm sur}) = \frac{1}{8\pi  G} \int_{\mathcal{V}} d^4 x\,  \sqrt{-g}\, \nabla_a (R^a_b \xi^b)
\label{basic1}
\end{equation}

The integration of the divergences in Eqs.(\ref{delmat}),(\ref{basic1}) lead to surface terms which contribute only on the horizon, since $\xi^a$ is nonzero only on the horizon. Further, since $\xi^a$ is in the direction of the normal,
 the demand $\delta A_{tot}=0$ leads to the result $(R^a_b-8\pi GT^a_b)\xi^b\xi_a=0$. Since $\xi^a$ is arbitrary except for the fact that it is null, this requires  $R^a_b-8\pi GT^a_b=F(g)\delta^a_b$, where $F$ is an arbitrary function of the metric. But since $\nabla_a T^a_b=0$ identically, $R^a_b-F(g)\delta^a_b$ must have zero divergence; it follows that $F$ must have the form
$F=(1/2)R+\Lambda$ where $R$ is the scalar curvature and $\Lambda$ is an undetermined (alas!) cosmological constant.
The resulting equation is
\begin{equation}
R^a_b-(1/2)R\delta^a_b+\Lambda\delta^a_b=8\pi GT^a_b
\end{equation}
which is identical to Einstein's equation. \textit{Nowhere did we need the bulk term in Einstein's action!}

We believe this derivation brings us closer to understanding the true nature of gravity. Since $(-A_{sur})$ is the entropy,
 its variation, when the horizon is infinitesimally moved, is equivalent to the change in the entropy $dS$ due to virtual work. The 
 variation of the matter term contriutes the $PdV$ and $dE$ terms and the entire variational principle is equivalent to the thermodynamic identity $TdS=dE+PdV$ applied to the changes when a horizon undergoes a virtual displacement. In the cae of spherically symmetric spacetimes, for example, this can
explicitly worked out\cite{pr75}. Thus \textit{Einstein's equations can be interpreted as the thermodynamic limit of microscopic,
statistical mechanics of `atoms of spacetime'} the structure of which we do not know (This approach has a long history\cite{sakharov}
but our result gives it a different, precise and elegant characterisation).

This approach is also logically coherent: Principle of Equivalence implies gravity affects light rays and thus affects causal structure; this leads to existence of observers who has access only to part of spacetime; this, in turn, forces us to having boundary terms in action the form of which can be determined by general considerations. When the causal horizon of the observers is interpreted in the 'membrane paradigm', one is led to consider the virtual work done by its displacements. This relation, which is in the form of $TdS=dE+PdV$, is identical to Einstein's equations. 

This approach suggests that the relevant degrees of freedom of gravity for a volume $\mathcal{V}$
reside in its boundary $\partial\mathcal{V}$, making gravity intrinsically holographic. (This result is also borne out by a study of 
Hilbert action in the Riemann normal coordinates; the $\Gamma^2$ part vanishes and the full contribution arises from the total divergence term).
There are obvious implications for quantum gravity and path integral formulation which require further study.

 Incidentally,  Eq.~(\ref{basic1}) shows that $\delta A_{\rm sur}=0$ (i) in all vacuum spacetimes with $R^a_b=0$, generalizing the previous result that $\partial_a \delta P^a=0$ in flat
 spacetime, or (ii) when $\xi^a =0$ on $\partial \mathcal{V}$ which is the usual
 textbook case. (This is why $\delta A_{\rm bulk}=0$ gives covariant field equations even though
 $A_{\rm bulk}$ is not generally covariant.)

\section{Derivation of  Einstein-Hilbert action}

The total Lagrangian for gravity in Eq.~(\ref{firsteqn}) is now $[L_{bulk}\sqrt{-g}-\partial_aP^a/16\pi G]$ which depends on the second derivatives of the metric through the $\partial_aP^a$ term. Such lagrangians, having second derivatives (which does not affect the equation of motion), have a natural interpretation in terms of the momentum space representation. To see this, recall that the quantum amplitude for 
for the dynamical
variables to change from $q_1 $ (at $t_1$) to $q_2$ (at $t_2$) is given by
\begin{equation}
K \left( q_2,t_2;q_1,t_1 \right) = \sum\limits_{\rm paths}
\exp \left[ \frac{i}{\hbar} \int dt~L_q(q,\dot q) \right] ~,
\label{qsopa}
\end{equation}
where the sum is over all paths connecting $(q_1,t_1)$ and $(q_2,t_2)$,
and the Lagrangian $L_q(q,\dot q)$ depends \textit{only} on $(q,\dot q)$. 
When we study the same system in momentum space, we need to determine
the corresponding amplitude for the system
to have a momentum $p_1$ at $t_1$ and $p_2$ at $t_2$, which 
 is given by the Fourier transform
of $K \left( q_2,t_2;q_1,t_1 \right)$ on $q_1,q_2$. The path integral representation of this momentum space amplitude is:
\begin{eqnarray}
&&\!\!\!\!\!\!\mathcal{G} \left( p_2,t_2;p_1,t_1 \right)\nonumber \\
&&\!\!\quad = \sum\limits_{\rm paths}
\int dq_1 dq_2 \exp \left[ \frac{i}{\hbar}
\left\{ \int dt~L_q - \left( p_2 q_2 - p_1 q_1 \right) \right\} \right]
\nonumber \\
&&\!\!\quad =\sum\limits_{\rm paths} \int dq_1 dq_2 \exp \left[ \frac{i}{\hbar}
\int dt \left\{ L_q - \frac{d }{ dt} \left( pq \right) \right \} \right]
\nonumber \\
&&\!\!\quad \equiv \sum_{\rm paths} {} \exp \left[ \frac{i}{\hbar}
\int L_p(q, \dot q, \ddot q)~dt \right] ~.
\label{lp}
\end{eqnarray}
where
\begin{equation}
L_p \equiv L_q - \frac{d}{ dt} \left( q\frac{\partial L_q }{ \partial\dot q} \right) ~.
\label{lbtp}
\end{equation}
In arriving at the last line of Eq.~(\ref{lp}), we have
redefined the sum over paths to include integration over $q_1$ and $q_2$.
This result shows that, given any Lagrangian $L_q(q,\partial q)$
involving only up to the first derivatives of the dynamical variables,
it is \emph{always} possible to construct another Lagrangian
$L_p(q,\partial q,\partial^2q)$ involving up to second derivatives, by Eq.~(\ref{lbtp})
such that it describes the same dynamics but with different boundary
conditions.
While using $L_p$, one keeps the \emph{momenta}  fixed
at the endpoints rather than the \emph{coordinates}.
 
  Thus, in the case of gravity,  the {\it same}  equations
    of motion can be obtained from $A_{\rm bulk}$ or from another  action:
   \begin{equation}
    \label{aeh}
   A_{grav} = \int d^4x \sqrt{-g} L_{\rm bulk} - \int d^4x \partial_c \left[ g_{ab}
    \frac{\partial \sqrt{-g} L_{\rm bulk} }{\partial(\partial_c g_{ab})}
    \right] 
 \end{equation}       
We  can now identify the second term in Eq.~(\ref{aeh})
with the $A_{sur}$ in Eq.~(\ref{sur}) thereby obtaining an equation that determines
 $L_{\rm bulk}$:
 \begin{equation}
 \left(\frac{\partial \sqrt{-g}L_{\rm bulk}}{\partial g_{ab,c}}g_{ab}\right)=
 P^c= - \frac{1}{ 16\pi G}\frac{1}{\sqrt{-g}} \partial_b(gg^{bc})
 \label{dseq}
\end{equation}
    It is straightforward to show that this equation is satisfied by the Lagrangian
\begin{equation}
\sqrt{-g}L_{\rm bulk}  = 
 \frac{\sqrt{-g} \, g^{ik}}{ 16\pi G} \left(\Gamma^m_{i\ell}\Gamma^\ell_{km} -
\Gamma^\ell_{ik} \Gamma^m_{\ell m}\right).
\label{ds}
\end{equation}  
    This Lagrangian  is precisely
    the first order Dirac-Schrodinger Lagrangian for gravity (usually called the $\Gamma^2$
    Lagrangian).
    Given the two pieces $\sqrt{-g}L_{\rm bulk}$ and $-\partial_a P^a$, the final second order Lagrangian  is, of course, just the sum, which turns out to be the standard Einstein-Hilbert Lagrangian:    
   \begin{equation}
  \sqrt{-g} L_{grav}=\sqrt{-g}L_{\rm bulk} - \partial_c P^c =  \left(\frac{R\sqrt{-g}}{ 16\pi G}\right).
      \label{lgrav}
       \end{equation}

Since Eq.(\ref{dseq}) involves only $\partial L_{\rm bulk}/\partial(\partial_a g_{bc})$, one can add a constant to $L_{\rm bulk}$
without affecting anything. Observations suggest that our universe has a cosmological constant (or something
which acts very similar to it \cite{ccetc}) and --- unfortunately --- our argument does not throw any light on this vital issue.
We shall comment on this aspect towards the end.

\section{Structure of Hilbert action}        
   
The first striking feature of Hilbert action is that it contains the second derivatives of the dynamical variables and hence a surface term. This feature had {\it no} explanation in conventional
approaches, while it arises most naturally in the derivation given above.

Second, and probably \textit{the most vital point}, is that the bulk and surface terms of the Hilbert action are related to each other by a very definite relation, viz., Eq.(\ref{aeh}). Not only that this relation has no explanation in conventional approaches, it has not even been noticed or discussed
in standard text books \cite{comment}. In the current approach, this relation again arises naturally and is central to determining the bulk term from the surface term.

Closely related to this is the third fact that neither $L_{bulk}$ nor $L_{sur}$ is geometrical. This is obvious in the Euclidean sector, in which we allowed for an extra vector field (the normal to the boundary) in the action, in addition to the metric tensor. It is central to our philosophy that the terms in the action can be (and indeed will be) different for different families of observers, since they will have access to different regions of spacetime. In fact, existence of a horizon is {\it always} dependent on the family of observers we consider. Even in Schwarzschild spacetime, in which a purely geometrical definition \textit{can} be given to event horizon, the observers who are freely falling into the black hole will have access to more information than the observers who are stationary on the outside. Thus we expect the action to be 
foliation dependent though generally covariant. 

The difference between foliation dependence and general covariance is worth emphasizing: One would have considered a component
of a tensor, say, $T_{00}$ as not generally covariant. But a quantity $\rho=T_{ab}u^au^b$ is a generally covariant scalar which will reduce to $T_{00}$ in a local frame in which $u^a=(1,0,0,0)$.
It is appropriate to say that $\rho$ is generally covariant but foliation dependent. In fact, any term which is not generally covariant can be recast in  a generally covariant form by introducing a foliation dependence. The surface term  and the bulk term  we have obtained are foliation dependent and will be different for different observers. But the full action  is, of course, foliation independent (We will see below that the full action can be interpreted as the thermodynamic free energy of spacetime). Incidentally, the content of Einstein's equations can be
stated entirely in terms of foliation dependent \textit{scalar} quantities as follows: The scalar projection of $R_{abcd}$ orthogonal to the vector field $u^a$ is $16\pi G\rho$
for all congruences. The projection is $R_{abcd}h^{ac}h^{bd}=2G_{ac}u^au^c$ where $h^{ac}=(g^{ac}+u^au^c)$
and the rest follows trivially;
this is a far simpler statement than the one found in some text books \cite{mtw}.

\subsection{Action is the Free Energy of Spacetime}

I will now turn to the form of 
 the Einstein-Hilbert action  for
stationary spacetimes and show that, in this case it has a natural decomposition of the action into
energy and entropy terms, the latter being a surface integral. Such spacetimes have a timelike Killing vector $\xi^a$ in (at least) part of the region. While the discussion below can be done in a manifestedly covariant manner, it is clearer to work in a frame in which $\xi^a=(1,0,0,0)$.
In any stationary spacetime, the $R^a_0$ components, in particular $R^0_0$, can be expressed as a divergence term 
\begin{equation}
R^0_0=\frac{1}{\sqrt{-g}}\partial_\alpha(\sqrt{-g}g^{0k}\Gamma^\alpha_{0k})
\label{roo}
\end{equation}
This is most easily seen from the identity for the Killing vector 
\begin{equation}
R^a_j\xi^j=R^a_0=\nabla_b\nabla^a\xi^b=\frac{1}{\sqrt{-g}}\partial_b(\sqrt{-g}\nabla^a\xi^b)
\end{equation}
where the last relation follows from the fact that $\nabla^a\xi^b$ is an antisymmetric tensor.
Eq.(\ref{roo}) now follows directly on noticing that all quantities are time-independent.
(For an index-gymnastics proof of the same relation, see \cite{ll2} section 105).
On the other hand, in any spacetime, one has the result:
$-2G^0_0=-16\pi{\cal H}_{ADM}=16\pi\rho$
where the last relation holds on the mass shell. We can now express the Einstein-Hilbert action as an integral
over $R=2(R^0_0-G^0_0)$. Since the spacetime is stationary, the integrand is independent of time and we need to limit
the time integration to a finite range $(0,\beta)$ to get a finite result. Converting the volume integral of $R^0_0$ over 3-space
to a surface integral over the 2-dimensional boundary,  we can write the Einstein-Hilbert action in any stationary spacetime as
\begin{eqnarray}
A_{\rm EH}
&=&\beta\int N\sqrt{h} d^3x\; \rho +\frac{\beta}{8\pi} \int  d^2x\sqrt{\sigma}\; 
Nn_\alpha(g^{0k}\Gamma^\alpha_{0k})\nonumber\\
&\equiv&\beta E-S
\label{two}
\end{eqnarray}
where  $N=\sqrt{-g_{00}}$ is the lapse function, $h$ is the determinant of the spatial metric
and $\sigma$ is the determinant of the 2-metric on the surface. 
In a class of  stationary metrics with a horizon and associated temperature the time interval has natural 
 periodicity in $\beta$, which can be identified with the inverse temperature. Also note that the $\beta N$ factor in Eq.(\ref{two}) 
 is again exactly what is needed to give the local Tolman temperature $T_{loc}=\beta^{-1}_{loc}\equiv(\beta N)^{-1}=T/\sqrt{-g_{00}}$ so
 one is actually integrating $\beta_{loc}\rho$ over all space, as one should, in defining $E$. When the 2-surface is a horizon,
 the integral over $R^0_0$ gives the standard expression for entropy obtained earlier.  This allows identification of the two terms with energy and entropy; together the Einstein action can be interpreted as giving the free energy of space time.

It is possible to obtain the above decomposition more formally.
The curvature tensor
 $R^{ab}_{cd}$ has a natural decomposition in terms of 3 spatial
tensors (see sec. 92 of \cite{ll2})  corresponding to 
$
{\cal S}^\alpha_\beta=R^{0\alpha}_{0\beta};
{\cal E}^\alpha_\beta
=(1/4)\epsilon^\alpha_{\phantom{\alpha}\mu\nu}R^{\mu\nu}_{\rho\sigma}\epsilon_\beta^{\phantom{\beta}\rho\sigma}$ (and 
$ B^\alpha_\beta=\epsilon_\beta^{\phantom{\beta}\mu\nu}R^{0\alpha}_{\mu\nu}$ which we will
not need).
In {\it  any spacetime}, the trace of these tensors have a simple physical meanings:
$2{\cal E}^\alpha_\alpha=-2G^0_0=16\pi\rho$
is essentially the numerical value of ADM Hamiltonian density while 
${\cal S}\equiv {\cal S}_\alpha^\alpha=R^0_0$. Since the scalar curvature is
$R=2({\cal S}+ {\cal E})$ we can make the decomposition:
\begin{equation}
A_{grav}=\frac{\beta}{8\pi}\int N\sqrt{h} d^3x\ [{\cal S}+ {\cal E}]=-S+\beta E
\label{one}
\end{equation}
in any stationary spacetime.
This shows that ${\cal S}$ and ${\cal E}$ are true scalars in 3-dimensional subspace and the decomposition has a geometric significance. 

In fact, one can go further and provide a 4-dimensional covariant (but foliation dependent)
description of the entropy and energy of spacetime. Given a foliation based on a timelike congruence of observers
(with $u^a$ denoting the four velocity) one can write:
$
{\cal S}^a_b=R^{ia}_{jb}u_iu^j;
{\cal E}^a_b={}^*R^{ia}_{jb}u_iu^j
$
where $^*R^{ab}_{cd}$ is the dual of the curvature tensor: $^*R^{ab}_{cd}
=(1/4)\epsilon^{ab}_{\phantom{ab}mn}R^{mn}_{rs}\epsilon_{cd}^{\phantom{cd}rs}$. 
It is easy to see that the contraction of ${\cal S}^a_b,{\cal E}^a_b$ with $u^a$ on any of the indices vanishes, so that they are essentially ``spatial" tensors; clearly, they reduce to the previous definitions when $u_0=(1,0,0,0)$.
The action can now be written with a covariant separation of the two terms:
\begin{equation}
A_{grav}=\frac{1}{8\pi}\int \sqrt{-g} d^4x\ {\rm Tr}[(R^{ia}_{jb}+{}^*R^{ia}_{jb})u_iu^j]
\label{oneone}
\end{equation}
where the trace is over the remaining indices. Given a family of observers with four velocities
$u^i$ this equation identifies an energy and entropy perceived by them.
This approach has a direct geometrical significance, since the representation of the curvature 
as differential form $\mathcal{R}^{ab}=\textbf{d}\omega^{ab}+\omega^a_{\phantom{a}c}\wedge\omega^{cb}
=R^{ab}_{|cd|}\omega^c\wedge\omega^d$ (see sec 14.5, ex. 14.14 of \cite{mtw}) uses a matrix representation with ${\cal S}^\alpha_\beta,{\cal E}^\alpha_\beta$ as the block diagonal terms and
with the action becoming the trace.

As an explicit example, consider the class of  stationary metrics, parametrized by a vector field $\textbf{v}(\textbf{x})$
is given by
\begin{equation}
ds^2=-dt^2+(d\textbf{x} -\textbf{v} dt)^2
\label{flow}
\end{equation} 
Consider first the spherically symmetric case in which $\textbf{v}=v(r)\hat{\textbf{r}}$
with $v^2\equiv-2\phi$. In this case, $2{\cal S}=-2\nabla^2\phi,2{\cal E}=-(4/r^2)(r\phi)'$. 
Let us compute the contribution of the action on the horizon at $r=a$, where $v^2=1,-\phi'=\kappa$.
Evaluating the integral over $R=2({\cal S}+{\cal E})$,  we get the contribution:
\begin{equation}
A_{grav}=-\pi a^2+\beta\frac{a}{2}=-S+\beta E
\end{equation} 
The first term (one quarter of horizon area, which arises from $2{\cal S}$) is the entropy and we can interpret the second term 
(which arises from $2{\cal E}$) as $\beta E$. This becomes  $\beta M$ in the case of Schwarzschild metric \cite{bj}. Hence the full action has the interpretation of
$\beta F$ where $F$ is the free energy. (See \cite{pr75,pr161} for more details). Thus the extremisation of the action corresponds to extremising the free energy, thereby providing a fully thermodynamic interpretation. (Incidentally, for the Schwarzschild metric $S=4\pi M^2,\beta E=8\pi M^2$ making
$S-\beta E=-4\pi M^2=-S$ ! The temptation to interpret the full action as entropy should be resisted,
since this is a peculiar feature special to Schwarzschild). 

Similar results arise for all metrics in Eq.(\ref{flow}). 
These metrics \cite{pg} have the following properties: (i) The spatial ($dt=0$) sections are flat
with $^3R=0$. (ii) The metric has unit determinant in Cartesian spatial coordinates.(iii)
The acceleration field $a^i=u^j\nabla_j u^i$ vanishes.  (iv) There is a horizon on the surface $v^2(\textbf{x})=1$. (iv) The extrinsic curvature is $K_{\alpha\beta}=(1/2)(\partial_\alpha v_\beta 
+\partial_\beta v_\alpha)$. (v)
For this spacetime, 
$2{\cal E}$ again gives the energy density; the entropy term is:
 $2{\cal S}=\nabla\cdot[-\nabla\phi+(\textbf{v}\cdot\nabla)\textbf{v}]$. The integral now gives the normal component
 of $\textbf{a}=-\nabla\phi+(\textbf{v}\cdot\nabla)\textbf{v}$ which has two contributions to the acceleration: the
 $-\nabla\phi$ (with $v^2=-2\phi$) is the standard force term while $(\textbf{v}\cdot\nabla)\textbf{v}$ is the "fluid"
 acceleration $d\textbf{v}/dt$ when $\partial \textbf{v}/\partial t=0$. So the interpretation of surface gravity leading to entropy holds
 true in a natural manner. 
 
 If $\textbf{v}$ is irrotational, then the two terms are equal and we get $2{\cal S}
 =-2\nabla^2\phi$ leading to an entropy which is one-quarter of the area of the horizon. Thus the result for spherically symmetric case holds for all metrics of the form in Eq.(\ref{flow}) with
 $\nabla\times \textbf{v}=0$.

 \subsection{There is more to gravity than gravitons}

Another striking feature of Einstein-Hilbert action --- again, not emphasized in the literature --- is that it is \textit{non analytic} in the coupling constant when perturbed around flat spacetime. To see this, consider the expansion of the action in terms of a ``graviton field" $h_{ab}$, by $g_{ab}=\eta_{ab}+\lambda h_{ab}$
where $\lambda= \sqrt{16\pi G} $ has the dimension of length and $h_{ab}$ has the correct dimension of
(length)$^{-1}$ in natural units with $\hbar=c=1$.  Since the scalar curvature has the structure $R\simeq (\partial g)^2+\partial^2g$, substitution of $g_{ab}=\eta_{ab}+\lambda h_{ab}$ gives to the lowest order:
\begin{equation}
L_{EH}\propto \frac{1}{\lambda^2}R\simeq (\partial h)^2+\frac{1}{\lambda}\partial^2h
\end{equation}
Thus the full Einstein-Hilbert Lagrangian is non-analytic in $\lambda$ because the \textit{surface term} is 
non-analytic in $\lambda$! It is sometimes claimed in literature that one can
obtain Einstein-Hilbert action for gravity by starting with a massless spin-2 field
 $h_{ab}$ coupled to the energy momentum tensor $T_{ab}$ of other matter sources to the lowest 
 order,  introducing self-coupling of $h_{ab}$ to its own energy momentum tensor at the
 next order 
 and iterating the process.  
It will be preposterous if, starting from the Lagrangian for the spin-2 field, $(\partial h)^2$, and doing a honest iteration on $\lambda$, one can obtain a piece which is \textit{non-analytic} in $\lambda$ (for a detailed discussion of this and related issues, see \cite{pr156}). At best, one can hope to get the quadratic part of $L_{EH}$ which gives rise to the $\Gamma^2$ action $A_{\rm bulk}$ but not the four-divergence term involving $\partial^2g$. The non-analytic nature of the surface term is vital for it to give a finite contribution on the horizon and the horizon entropy cannot be interpreted in terms of gravitons propagating around Minkowski spacetime. Clearly, there is lot more to gravity than gravitons.

This result has implications for the $G\to0$ limit of the  entropy term $S=(1/4)(\mathcal{A}_\perp/G)$ we have obtained. If the transverse area  $\mathcal{A}_\perp$ scales as $(GE)^2$ where $E$ is an energy scale in the problem (as in Schwarzschild geometry), then 
$S\to 0$ when $G\to0$. On the other hand if $\mathcal{A}_\perp$ is independent of $G$ as in the case of e.g., De Sitter universe with
$\mathcal{A}_\perp=3\pi/\Lambda$, where $\Lambda$ is an independent cosmological constant in the theory, \textit{unrelated} to $G$ (or
even in the case of Rindler spacetime) then the entropy diverges as $G\to0$. (In all cases, the entropy diverges as $\hbar\to0$).
This non analytic behaviour for a term in action, especially in the case of flat spacetime in nontrivial coordinates, is reminiscent of the $\theta$ vacua in gauge theory. We shall explore this feature more closely in the next section.

\section{Singular coordinate transformations and non trivial topology}

The Euclidean structure of a wide class of spacetimes with horizon
is correctly represented by the Euclidean Rindler metric in Eq.~(\ref{eext}). This
is obvious from the fact that the horizon is mapped to the origin of the $N-\tau$
plane which is well localized in the Euclidean sector, making the approximation by
a Rindler metric rigorously valid. It is therefore important to understand how
non trivial effects can arise ``just because'' of a coordinate
transformation from the inertial to Rindler coordinates. After all, in the inertial coordinates in flat spacetime $\Gamma^a_{bc}=0$ making both $L_{bulk}$ and $P^a$ individually zero while in the Rindler coordinates $P^a\ne 0$ leading to Eq.(\ref{seven}). We shall provide
some insights into this issue.

We begin by recalling some formal analogy between gravity and non-Abelian gauge theories. If
the connection coefficients $\Gamma^j_{ak}\equiv (\Gamma_a)^j_k$ are represented as the elements of  matrices
$\Gamma_a$ (analogous to the the gauge potential $A_a$), then the curvature tensor can be represented as
\begin{equation}
R_{ab}=\partial_a\Gamma_b-\partial_b\Gamma_a +\Gamma_a\Gamma_b-\Gamma_b\Gamma_a
\end{equation}   
(with two matrix indices suppressed) in a form analogous to the gauge field $F_{ab}$.
Consider now an infinitesimal coordinate transformation $x^a\to x^a+\xi^a$ from a local inertial frame to an accelerated frame.
The curvature changes by 
\begin{equation}
\delta R_{ab}=\partial_a\delta\Gamma_b-\partial_b\delta\Gamma_a 
\label{deltarab}
\end{equation} 
since $\Gamma\delta\Gamma$ term vanishes in the local inertial frame. For a coordinate transformation in flat spacetime, we will expect
 $\delta\Gamma_a$ to be pure gauge in the form $\delta\Gamma_a=\partial_a\Omega$, so that
 $\delta R_{ab}=0$. For $x^a\to x^a+\xi^a$, we do have $\delta\Gamma_a=\partial_a\Omega$ where $\Omega$ is a matrix with
elements $\Omega^i_j=-\partial_j\xi^i$ so it would seem that $\delta R_{ab}=0$. However, there is subtlety here.

Recall that, in standard flat spacetime electrodynamics one can have  vector potential $A_i = \partial_i q(x^a)$ which appears to be a pure gauge connection but can have non-zero field strengths.  If we take $x^a =(t, r, \theta, \phi)$ and
$q(x^a) = \phi$, then the vector potential $A_i = \partial_i(\phi)$ is \emph{not}
pure gauge and will correspond to a magnetic flux confined to an Aharanov-Bohm type solenoid
at the origin. This is easily verified from noting that the line integral of $A_idx^i$ around 
 the origin  will lead to a non-zero result, showing
 $\nabla \times {\bf A}$ is non zero
at the origin corresponding to $x^2 + y^2 =0$ in the Cartesian coordinates. In this case $q(x^a) =\phi$ is a periodic coordinate
which is the reason for the nontrivial result.

The same effect arises in the the case of a transformation from inertial to Rindler coordinates near any horizon
in the \textit{Euclidean} sector in which $\tau$ is periodic. In this case, $\xi^a=(-\kappa t x, -(1/2)\kappa t^2,0,0)$ and the matrix
$\Omega^i_j=-\partial_j\xi^i$ has the nonzero components $\Omega^0_0=\kappa x,\Omega^1_0=\kappa t=\Omega^0_1$. The line integral
of $\delta\Gamma_0$ over a circle of radius $x$ around the origin in Euclidean $\tau-x$ plane is given by
\begin{equation}
\oint\delta\Gamma_0 dx^0=\int_0^{2\pi/\kappa}d\tau\partial_0\Omega=\Omega\bigg\vert_0^{2\pi/\kappa}=2\pi
\label{gammaint}
\end{equation} 
for the $(^0_1),(^1_0)$ components of the matrix $\Omega$.
The nonzero value for this integral shows that the curvature $\delta R_{ab}$ in Eq.(\ref{deltarab}) is nonzero and is concentrated on the origin of the $\tau-x$ plane (which is the horizon in Lorentzian spacetime)!
Just as the Aharanov-Bohm
    effect introduces non zero winding number in space, the concentrating of the scalar curvature
    in the origin of the Euclidean plane leads to a non zero winding number in the presence of 
    horizons and thermal effects. 
    Since the topological feature arises due to a circle of infinitesimal radius around the origin in the 
    Euclidean case, the analysis should work for any horizon which can be approximated by a Rindler
    metric near the horizon.

In  non-Abelian gauge theories there exist pure gauge configurations in which
the gauge potential cannot be made to vanish everywhere. These configurations, usually called
theta vacua, exist because of the existence of gauge transformations which cannot
be continuously deformed to identity. In the case of gravity, the analogy is provided
by \textit{singular} coordinate transformation which converts a line 
element with $g_{00} <0$ everywhere to a metric with $g_{00}=-N^2({\bf x})$    vanishing
at some surface.  The transformation from inertial coordinates to Rindler coordinates
or, more generally, from the Kruskal type coordinates to Schwarzschild type coordinates
belongs to this class. An immediate effect of such a transformation is that analytic continuation in $t$
makes $\tau=it$ a periodic (``angular") coordinate, with period $2\pi/\kappa$ leading to the result in Eq.(\ref{deltarab}).

A new issue, which is conceptually important, arises while doing quantum field theory in  a spacetime
  with a $N=0$ surface.
  All physically relevant results in the spacetime will depend on the 
  combination $Ndt$ rather than on the coordinate time $dt$. The Euclidean rotation $t\to t e^{i\pi/2}$
  can equivalently be thought of as the rotation $N \to N e^{i\pi/2}$. This procedure 
  becomes ambiguous on the horizon at which $N=0$. But the family of observers with a horizon, {\it will}
indeed be using a comoving co-ordinate system in which $N\to 0$ on the horizon.
 Clearly we need a new physical principle to handle quantum field theory as seen by this family of observers.

 One possible way is to regularize $g_{00}$ and treat the Rindler type metric 
 as a limit of a sequence of metrics parameterized by a regulator $\epsilon$.
 The nature of the regulator can be obtained by noting that 
 the Euclidean rotation is equivalent to the $i\epsilon$ prescription in which one uses the transformation
  $t \to t (1+i\epsilon)$ which, in turn, translates to $N\to N(1+i\epsilon)$. Expanding this out, we get 
  $  N \to N + i \epsilon \ {\rm sign}(N)$, which can be combined into the form $N^2 \to N^2 + \epsilon^2$. To see the effect of this regulator, let us consider \cite{pr59} a class of metrics of the form
 \begin{equation}
ds^2 = -f (x) dt^2 + dx^2 + dy^2+dz^2
\label{startmetricone}
\end{equation}
where (i)$f$ is an even function of $x$, (ii)$f>0$ for all $x$ and  (iii)the metric is asymptotically Rindler: $f(x^2) \to \kappa^2 x^2 $ as $ x^2 \to \infty)$.  The Einstein Hilbert action for these metrics is
  \begin{equation}
A  =-2\left( \frac{\mathcal{A}_\perp}{4G}\right) \left(\frac{\kappa\beta}{2\pi}\right) 
\label{actionval}
\end{equation} 
Since $R$ is independent of $t$ and the transverse coordinates, we have to restrict the integration
over these to a finite range with $0\le t \le \beta$ and $\mathcal{A}_\perp$ being the transverse area. We see that the result is
completely independent of the detailed behaviour of $f(x)$ at finite $x$. 
Let us now consider the class of two parameter metrics with $f(x)=\epsilon^2+\kappa^2x^2$. 
 When $\kappa=0$ this metric represents flat spacetime in  standard Minkowski coordinates and the action in Eq.~(\ref{actionval}) vanishes; the metric  also represents flat spacetime for $\epsilon=0$ but now in the Rindler coordinates. 
 Since the result in Eq.~(\ref{actionval})
holds independent of $\epsilon$, it will continue to hold even when we take the limit of $\epsilon$ tending to zero. But when $\epsilon$
goes to zero, the metric reduces to standard Rindler metric and one would have expected the scalar curvature to vanish identically, making $A$ vanish identically! Our result in (\ref{actionval})    shows that the action is finite even for a Rindler spacetime  \emph{if we interpret it as arising from the limit of this class of metrics.}
It is obvious that, treated in this limiting fashion, as $\epsilon$ goes to zero $R$ should become a distribution in $x^2$ 
such that it is zero almost everywhere except at the origin and has a finite integral.
For finite values of $(\epsilon,\kappa)$ the spacetime is curved with only nontrivial component:
\begin{equation}
R = -\frac{2\epsilon^2 \kappa^2}{(\epsilon^2 + \kappa^2 x^2)^2} = - \frac{1}{2} R^t_{\phantom{t}x tx}
\label{reqn}
\end{equation}
There is no horizon when $\epsilon\ne 0$. When $\epsilon\ne 0, \kappa\to 0$ limit is taken, we obtain the flat spacetime in Minkowski coordinates without ever producing a horizon. But the limit $\kappa\ne 0,\epsilon\to0$ leads to a different result:
when $\epsilon\to 0$, a horizon appears at $x=0$ and, in fact,
the scalar curvature $R$ 
in Eq.~(\ref{reqn}) becomes   the distribution
\begin{equation}
\lim_{\epsilon\to 0} R =- 2 \delta(x^2) 
\label{rlimiteqn}
\end{equation}
showing that the curvature is concentrated on the surface $x^2=0$ giving a finite value to the action even though
the metric is almost everywhere flat in the limit of $\epsilon\to 0$.  The entire analysis goes through even in the Euclidean sector, showing that the curvature is concentrated on  $x^2=X^2+T_E^2=0$ which  agrees with the result obtained earlier in Eq.(\ref{gammaint}).

  \section{Elasticity of the spacetime solid}  

The analysis so far indicates a perspective towards gravity with the following key ingredients.

(i) The horizon perceived by a congruence of observers dictates the form of the action 
functional to be used by these observers. This action has a surface term which can
be interpreted as an entropy. 

(ii) The active version of the coordinate transformation $x^a \to x^a + \xi^a$ acquires
a dynamical content through 
our discussion in section III. Interpreting the virtual displacement of the horizon in terms of a 
thermodynamic relation, one can obtain the equations of gravity purely from the surface
term. 

(iii) The metric components become singular on the horizon in the coordinate 
system used by the congruence of observers who perceive the horizon. The transformation
from this coordinate system to a non singular coordinate system (like the locally
inertial coordinate system) near the horizon will require the use of a coordinate
transformation which itself is  singular. In the Euclidean sector, such transformations lead to
non trivial effects.

Given these results, it is interesting to pursue the analogy between spacetime and an
elastic solid further and see where it leads to. In particular, such an approach
will treat variables like $g_{ab}, T_{ab}, $ etc. as given functions which are not dynamic.
We should be able to obtain a variational principle in terms of some \textit{other} 
quantities but \textit{still} obtain Einstein's equations. I will now describe one such approach.

Let us begin by noting that, in
      the study of elastic deformation in continuum mechanics \cite{ll7}, one begins
      with the deformation field $u^\alpha (x) \equiv \bar x^\alpha -x^\alpha$ which indicates how 
      each point in a solid moves under a deformation.
       The deformation contributes to the thermodynamic functionals like free energy, entropy etc. In the absence
of  external fields, a constant $u^\alpha$ cannot make a contribution because of translational invariance. Hence,
to the lowest order, 
the thermodynamical functionals  will be quadratic in the scalars constructed from
    the derivatives of the deformation field.
       The derivative $\partial_\mu u_\nu$ can be decomposed into
      an anti symmetric part, symmetric traceless part and the trace corresponding to deformations
      which are rotations, shear and expansion. Since the overall rotation of the solid will  not  
     change the thermodynamical variables, only  the other two components, $S_{\mu\nu}
      \equiv \partial_\mu u_\nu +
      \partial_\nu u_\mu-(1/3)\delta_{\mu\nu}\partial_\alpha u^\alpha$ and $\partial_\alpha u^\alpha$,
      contribute.
      The extremisation of the relevant functional (entropy, free energy ....) 
 allows one to determine the equations which govern
      the elastic deformations.

      The analogue  of elastic deformations in the case of spacetime manifold will be the coordinate
      transformation $x^a \to \bar x^a = x^a+\xi^a(x)$.  
    Our paradigm requires us to take this transformation to be of fundamental importance rather than
      as ``mere coordinate relabelling''. In analogy with the elastic solid, we will attribute a
thermodynamic  functional --- which we shall take to be the entropy  ---
       with a given  spacetime deformation. This will be a quadratic functional of $Q_{ab} \equiv
      \nabla_a \xi_b$ in the absence of matter. The presence of matter will, however,  break the translational
      invariance and hence there could be a contribution which is quadratic in $\xi_a$ as
      well. We, therefore, take the form of the entropy functional to be  
      \begin{equation}
      S=\frac{1}{8\pi G}\int d^4x\, \sqrt{-g}\, \left[
      M^{abcd}   \nabla_a \xi_b  \nabla_c \xi_d + N_{ab} \xi^a\xi^b\right]
      \label{freeenergy}
      \end{equation}
      where the tensors, $M^{abcd}$ and $N_{ab}$ are yet to be determined.  They can depend on other
coarse grained macroscopic variables like the matter stress tensor $T_{ab}$, metric $g_{ab}$,  etc.  (The overall constant factor is again introduced with hindsight.)     
      Extremising $S$ with respect to the deformation field $\xi^a$ will lead to the equation
      \begin{equation}
      \nabla_a (M^{abcd}  \nabla_c)\xi_d = N^{bd} \xi_d
      \label{basic}
      \end{equation}
      In the case of elasticity, one would have used such  an equation to \textit{determine} the deformation
      field $\xi^a(x)$. 
       But the situation is quite different in the case of spacetime.  Here,
      in the  coarse grained limit  of continuum spacetime physics, one requires
      {\it any} deformation $\xi^a(x)$ to be allowed in the spacetime {\it provided
      the background spacetime satisfies Einstein's equations}. Hence,  if our ideas are correct,
      we should be able to choose $M^{abcd}$  and $N_{ab}$ in such a way that 
       Eq.(\ref{basic}) leads to Einstein's equation when we demand that it should hold for
       any $\xi^a(x)$. 
       
       Incredibly enough, this requirement is enough to uniquely determine
       the form of  $M^{abcd}$ and $N_{ab}$ to be: 
       \begin{equation}
       M^{abcd} = g^{ad}g^{bc} - g^{ab}g^{cd} ; N_{ab} = 8\pi G (T_{ab} - \frac{1}{2} g_{ab} T)
       \label{mndef}
       \end{equation}
       where $T_{ab}$ is the macroscopic stress-tensor of matter.
       In this case, the entropy functional becomes
        \begin{eqnarray}
      S&\propto&\int d^4x\, \sqrt{-g}\, \left[(\nabla_a \xi^b)(\nabla_b \xi^a) - (\nabla_b \xi^b)^2
       + N_{ab} \xi^a\xi^b\right]\nonumber\\
       &=& \frac{1}{8\pi G}\int d^4x \,  \sqrt{-g}\\
       &\times&  \left[ {\rm Tr}\ (Q^2) - ({\rm Tr}\ Q)^2 
       + 8\pi G \left(T_{ab} - \frac{1}{2} g_{ab} T\right)\xi^a \xi^b\right]\nonumber
       \label{tr}
      \end{eqnarray}
      where $Q_{ab}\equiv\nabla_a\xi_b$.
      The variation with respect to $\xi^a$ leads to the Eq.(\ref{basic}) which, on using Eq.(\ref{mndef}),
      gives:
      \begin{equation}
      (\nabla_a\nabla_b - \nabla_b\nabla_a) \xi^a = 8\pi G \left(T_{ab} - \frac{1}{2} g_{ab} T\right) \xi^a
      \end{equation}
      The left hand side is $R_{ab}\xi^a$ due to the standard identity for commuting the covariant
      derivatives. Hence the equation can hold for arbitrary $\xi^a$ only if
\begin{equation}
       R_{ab} =8\pi G \left(T_{ab} - \frac{1}{2} g_{ab} T\right)
\label{albie}
\end{equation}
      which is the same as Einstein's equations.  This  result is worth examining in detail:

To begin with, note that we did {\it not vary  the metric tensor} to obtain Eq.(\ref{albie}). In this approach,
$g_{ab}$ and $T_{ab}$ are derived macroscopic quantities and are \textit{not} fundamental variables. Einstein's equations arise as a consistency condition, reminiscent of the way it is derived in some string theory models due to the vanishing of beta function \cite{strings}. While the idea of spacetime being an ``elastic solid" has a long history (starting from 
Sakharov's work in \cite{sakharov}),
all the previous approaches try to obtain a low energy effective action in terms of $g_{ab}$s  and then vary $g_{ab}$  to get Einstein's equations. Our approach is {\it very} different and is a simple consequence of taking our paradigm seriously.

Second, this result offers a  new perspective on  general coordinate transformations which are  treated 
as akin to deformations in solids.  General covariance now arises as a macroscopic symmetry in the long wavelength limit, when the spacetime satisfies the Einstein's equations. In this limit,  the deformation should not change
the thermodynamical functionals.
      This is indeed true; the expression for the entropy in Eq.(\ref{freeenergy}) 
      reduces to a four-divergence   when Einstein's equations
      are satisfied (``on shell") making $S$  a surface term: 
      \begin{eqnarray}
      S &=& \frac{1}{8\pi G}\int_{\cal V} d^4x\, \sqrt{-g}\, \nabla_i ( \xi^b \nabla_b \xi^i - \xi^i \nabla_b \xi^b)\nonumber\\
      &=&\frac{1}{8\pi G}\int_{\partial{\cal V}} d^3x\, \sqrt{h}\, n_i( \xi^b \nabla_b \xi^i - \xi^i \nabla_b \xi^b)
\label{onsur}
      \end{eqnarray}
      {\it The entropy of a bulk region $\mathcal{V}$  of spacetime  resides in its boundary
      $\partial \mathcal{V}$ when Einstein's equations are satisfied.} In varying  Eq.(\ref{freeenergy})
      to obtain Eq.(\ref{basic}) we keep this surface contribution to be a constant. 

This result has an important consequence. If the spacetime has microscopic degrees of freedom,
then any bulk region will have an entropy and it has always been a surprise why the entropy scales as
the area rather than volume. Our analysis shows that, in the semiclassical limit, when Einstein's equations hold
to the lowest order, {\it the entropy is contributed only by the boundary term
and the system is holographic.}\cite{bj}

 This result can be connected with our earlier one in Eq.~(\ref{seven}) by  noticing that, 
      in the case of spacetime, there is one kind of ``deformation" which is rather special --- 
      the inevitable translation forward in time: $t\to t+\epsilon$. More formally, one can consider this as arising from $x^a\to x^a+\xi^a$ where $\xi^a=u^a$ is the unit normal to a spacelike hypersurface. 
      Then $\xi^a n_a =0$, making the second term in Eq.~(\ref{onsur}) vanish; the first term
 will lead to the integral over the surface gravity; in the Rindler limit it will give
      Eq.~(\ref{seven}).       
While the results agree, the interpretation is quite different. The deformation field corresponding to
time evolution hits a singularity on the horizon, which is analogous to a topological defect in a solid. The entropy is the price we pay for this defect. Alternatively, in the Euclidean sector,
the vector $u_i=\partial_it$ goes over to $\partial_i\theta$ where $\theta$ is a periodic
coordinate. The time translation becomes rotation around the singularity in the origin of $01$
plane.  

Finally, let us consider the implications of our result for the cosmological constant for which
$T_{ab}=\rho g_{ab}$ with a constant $\rho$. Then, $T_{ab} - \frac{1}{2} g_{ab} T=-\rho g_{ab}$
and the coupling term $N_{ab}\xi^a\xi^b$ for matter is proportional to $\xi^2$. If we vary 
Eq.(\ref{freeenergy}) but restrict ourselves to vectors of constant
      norm, then the vector field does not couple to the cosmological constant!
       In this case, one can show that the variation leads to the equation:
      \begin{equation}
      R_{ab} - \frac{1}{4} g_{ab} R=8\pi G(T_{ab} - \frac{1}{4} g_{ab} T)
      \label{unimod}
      \end{equation}
      in which both sides are trace free. Bianchi identity can now be used to show that
      $\partial_a(R+8\pi G T)=0$, requiring $(R+8\pi G T)=$ constant. Thus  cosmological constant arises
      as an (undetermined) integration constant in such models \cite{unimode}, and could be interpreted as a
      Lagrange multiplier that maintains the condition $\xi^2=$ constant.  This suggests that the effect of vacuum energy density
      is to rescale the length of $\xi^a$.  The quantum
    micro structure of spacetime at Planck scale is capable of readjusting itself, soaking up any
    vacuum energy density which is introduced. 
    Since this process  is inherently quantum gravitational,
    it is subject to quantum fluctuations at Planck scales. 
     The cosmological constant we measure corresponds to this  small 
    residual fluctuation  and will depend on the volume of the spacetime region that is probed.  
    It is small, in the sense that it has been reduced from $L_P^{-2}$ to $L_P^{-2}(L_PH_0)^2$, 
    (where $L_P$ is the Planck length and $H_0$ is the current Hubble constant)
    which indicates the fact that fluctuations --- when measured over a large volume --- is small
     compared to the bulk value.   A tentative implementation of this suggestion was made in
     ref.\cite{cqglambda} but one needs to work harder to find a microscopic realization of these ideas.    

     \section*{Acknowledgements}
     I thank Apoorva Patel and K.Subramanian for discussions.

\bibliography{apssamp}

\begin{thebibliography}{100}
\bibitem{comment0} The Greek indices vary over 1-3 and Latin indices go over 0-3.
Except for this, we use the conventions of ref.\cite{mtw}.
\bibitem{comment3}
I have not seen this expression in standard textbooks.  
One can obtain this directly, but tediously, from the textbook expressions for $g^{ab} \delta R_{ab}$; or, more cleverly, using the result that the \textit{local} variation $\delta (Q \sqrt{-g}) = -\sqrt{-g}\, \nabla_a (Q \xi^a)$ for any scalar $Q$. Simple manipulation with $Q=R$
gives the result.
\end{thebibliography}

\end{document}